\documentclass[preprint]{vgtc}                       
\graphicspath{{figures/}{pictures/}{images/}{./}} 

\usepackage{times}                     

\usepackage{tabu}                      
\usepackage{booktabs}                  
\usepackage{lipsum}                    
\usepackage{mwe}                       

\usepackage{mathptmx}                  
\usepackage{float}
\usepackage{amsmath}
\usepackage{enumitem}

\onlineid{0}

\vgtccategory{Research}

\vgtcinsertpkg

\preprinttext{%
    To appear in the Proceedings of the 19th IEEE Pacific Visualization Conference (PacificVis 2026), Sydney, Australia (Conference Track).
}

\title{InfoCIR: Multimedia Analysis for Composed Image Retrieval}

\author{Ioannis Dravilas\thanks{Contributed equally to this work.}%
\textsuperscript{\, ,\,}%
\thanks{Corresponding author: Ioannis Dravilas} %
\textsuperscript{,\,}%
\thanks{e-mail: \{ioannis.dravilas, ioannis.kapetangeorgis, anastasios.latsoudis\}@student.uva.nl} %
\and Ioannis Kapetangeorgis\footnotemark[1]\textsuperscript{\, ,\,}\footnotemark[3] %
\and Anastasios Latsoudis\footnotemark[1]\textsuperscript{\, ,\,}\footnotemark[3]
\and Conor McCarthy\thanks{e-mail: \{c.t.mccarthy, g.barretoferreiramarcelino, m.worring\}@uva.nl \\[0.05em] \textcopyright~2026 IEEE. Personal use of this material is permitted.  Permission from IEEE must be obtained for all other uses, in any current or future media, including reprinting/republishing this material for advertising or promotional purposes, creating new collective works, for resale or redistribution to servers or lists, or reuse of any copyrighted component of this work in other works.}
\and Gonçalo Marcelino\footnotemark[4]
\and Marcel Worring\footnotemark[4]}
\affiliation{\scriptsize University of Amsterdam, The Netherlands}

\teaser{
  \centering
  \includegraphics[width=\linewidth, alt={Alt}]{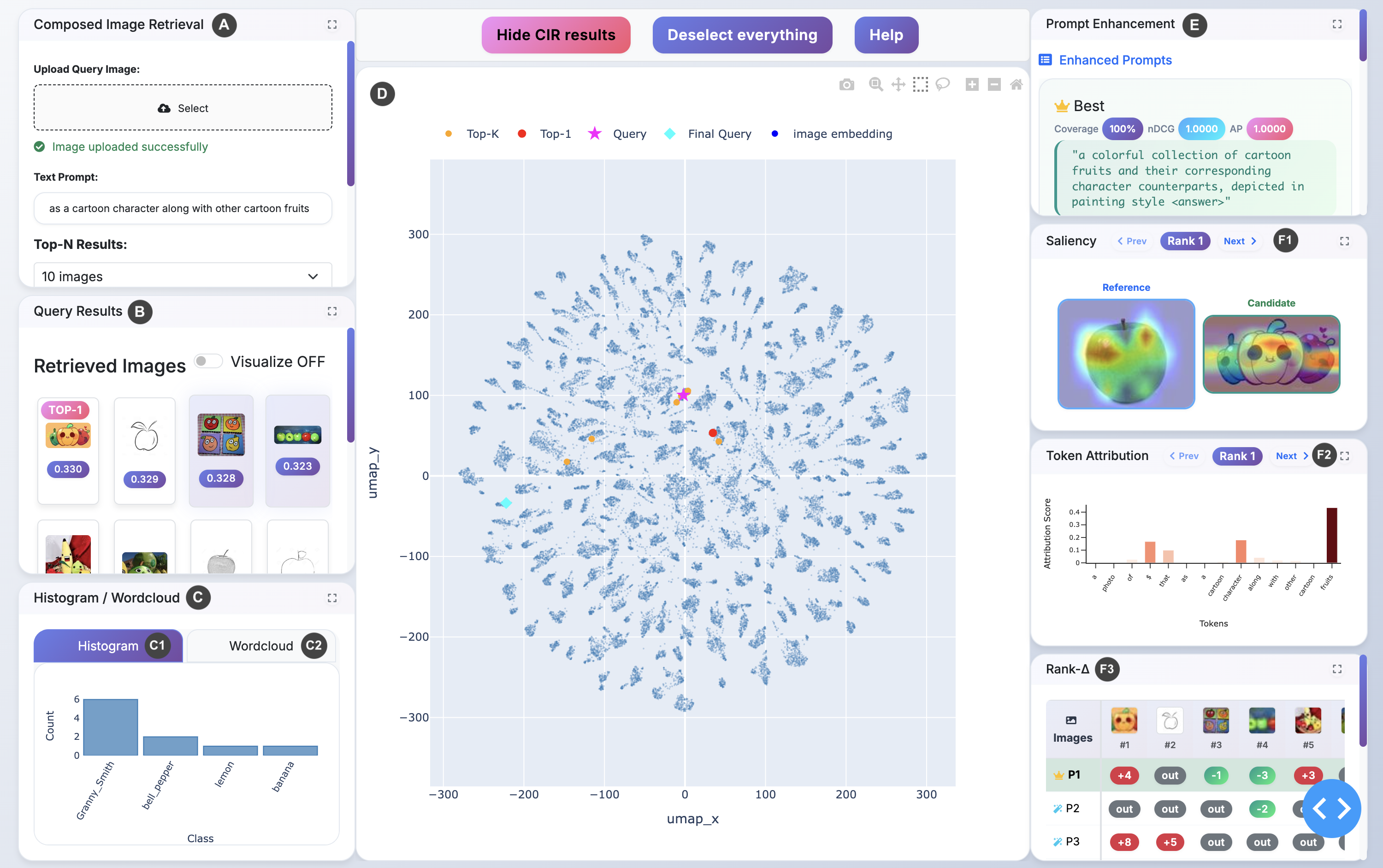}
  \caption{The user interface of our system consists of six main panels. (A) The Composed Image Retrieval Panel allows users to input an image and a text prompt to initiate the retrieval process and establish a baseline search. (B) The Query Results Panel displays the top-$k$ images which are listed by similarity; any image can be clicked to mark it as an ideal target for prompt refinement. (C) The Histogram / Word cloud Panel includes the class-frequency histogram (C$_1$) and a word cloud (C$_2$), summarizing labels within the current top-$k$. (D) The central Embedding View shows a 2D UMAP projection of the dataset, highlighting the reference image, the composed query embedding, and the top-$k$ results. (E) The Prompt Enhancement Panel proposes alternative prompts conditioned on the user-selected ideal target images, using an LLM, and retrieval metrics. (F) The Explanation Panel visualizes model attribution using a saliency map (F$_1$), a token attribution bar chart (F$_2$) and a Rank-$\Delta$ heatmap (F$_3$).}
  \label{fig:teaser}
}

\abstract{
    Composed Image Retrieval (CIR) allows users to search for images by combining a reference image with a text prompt that describes desired modifications. While vision-language models like CLIP have popularized this task by embedding multiple modalities into a joint space, developers still lack tools that reveal how these multimodal prompts interact with embedding spaces and why small wording changes can dramatically alter the results. We present InfoCIR, a visual analytics system that closes this gap by coupling retrieval, explainability, and prompt engineering in a single, interactive dashboard. InfoCIR integrates a state-of-the-art CIR back-end (SEARLE  \cite{agnolucci2024isearleimprovingtextualinversion}) with a six-panel interface that (i) lets users compose image + text queries, (ii) projects the top-$k$ results into a low-dimensional space using Uniform Manifold Approximation and Projection (UMAP) for spatial reasoning, (iii) overlays similarity-based saliency maps and gradient-derived token-attribution bars for local explanation, and (iv) employs an LLM-powered prompt enhancer that generates counterfactual variants and visualizes how these changes affect the ranking of user-selected target images. A modular architecture built on Plotly-Dash allows new models, datasets, and attribution methods to be plugged in with minimal effort. We argue that InfoCIR helps diagnose retrieval failures, guides prompt enhancement, and accelerates insight generation during model development. All source code allowing for a reproducible demo is available at \url{https://github.com/giannhskp/InfoCIR}.
} 


\begin{document}


\firstsection{Introduction}

\maketitle


Composed Image Retrieval (CIR) is the task of retrieving a ranked list of images that resemble a reference image but differ in a specific way described by a natural language prompt \cite{vo2018composingtextimageimage}. Unlike traditional image search, which typically relies on either visual similarity or textual queries alone, CIR fuses both modalities into a single query. For example, a user may ask for images similar to a photo of a forest, but ``in autumn colors'' or ``at night''. Solving this task requires the system to interpret the transformation described in the prompt and apply it to the visual concept. CIR has gained attention in multimedia retrieval and computer vision \cite{song2025comprehensivesurveycomposedimage, psomas2024composedimageretrievalremote}, especially with the introduction of vision-language models such as CLIP \cite{radford2021learningtransferablevisualmodels}, which embeds both images and text into a joint feature space. Joint image-text embeddings form the basis of many recent CIR models, including zero-shot systems that leverage pre-trained components without further task-specific training \cite{efthymiadis2024composedimageretrievaltrainingfree, baldrati2023zeroshotcomposedimageretrieval, LDRE}.

In practice, CIR remains difficult to work with. Existing workflows rely heavily on ``blind'' trial-and-error: developers diagnose a low Recall@K metric and randomly tweak the prompt or reference image without understanding the underlying failure mechanism. While general-purpose tools exist for embedding visualization \cite{smilkov2016embeddingprojectorinteractivevisualization} and unimodal explainability \cite{Selvaraju_2019}, they are insufficient for CIR for two distinct reasons. First, generic embedding projectors treat queries as static data points, failing to visualize the vector shift induced by the textual modifier, the core mechanism of CIR. Second, standard saliency maps often visualize visual features in isolation, failing to reveal when and why the retrieval model ignores the textual condition in favor of visual dominance (e.g., retrieving a ``green apple'' despite the prompt ``red''). Furthermore, as we observe in our analysis, standard dimensionality reduction techniques like UMAP can suffer from style bias in CLIP-based spaces, where images are clustered by artistic style rather than the semantic categories relevant to the user. These limitations hinder both system debugging and prompt engineering, especially in applications where precise control over image semantics is needed \cite{tang2025missingtargetrelevantinformationprediction}.

To address this gap, we present InfoCIR, a visual analytics system designed to support the exploration, diagnosis, and enhancement of CIR models. Developed following a design study methodology \cite{sedlmair2012design}, InfoCIR targets developers and researchers who face recurrent challenges in diagnosing retrieval failures and iteratively refining multimodal queries.

As illustrated in \cref{fig:teaser}, the interface integrates six linked components (A-F) that support interactive exploration. Users can select a reference image and a text prompt, view top-$k$ retrieval results, inspect their distribution in the embedding space through Uniform Manifold Approximation and Projection (UMAP) \cite{mcinnes2020umapuniformmanifoldapproximation}, and identify key factors that influence model behavior. The system overlays saliency maps, token attributions, and prompt variants to help users understand and refine the search process. See the caption of \cref{fig:teaser} for a detailed breakdown of each component.

We present InfoCIR explicitly as a system contribution, emphasizing completeness and practicality in bridging the gap between black-box retrieval models and human intent. Our goal is to operationalize retrieval, explainability, and prompt engineering into a coherent workflow that supports human-in-the-loop reasoning. Our specific contributions include:

\begin{itemize}
    \item A unified visual analytics architecture that integrates retrieval, attribution, and prompt engineering, enabling a ``diagnose-compare-enhance'' feedback loop to address the limitations of isolated, ``blind'' trial-and-error workflows.
    \item An interactive relevance-feedback workflow that couples LLM-generated prompt variants with visual verification via saliency maps, token-level attribution, and a novel Ideal-anchored Rank-$\Delta$ heatmap to verify prompt impact.
    \item A supervised metric projection pipeline utilizing a debiased UMAP approach to project results, reference images, and queries into a common space, specifically designed to overcome style bias for more accurate neighborhood analysis.
    \item A controlled user study demonstrating that InfoCIR significantly improves retrieval efficiency and success rates compared to a baseline CIR interface.
\end{itemize}

Together, these features turn InfoCIR into a practical tool for diagnosing retrieval failures, understanding model behavior, and refining CIR workflows in an interpretable and interactive manner.

\section{Related Work}
\label{ref:related_work}

This section reviews five research areas that shape the design and functionality of our system: (1) \textbf{Visual analytics for retrieval systems} provide interactive tools for examining ranking dynamics, analyzing query effects and refining results.
(2) \textbf{Dimensionality reduction and embedding visualization} project high-dimensional embeddings into navigable spaces, allowing users to inspect structural patterns and model behavior. 
(3) \textbf{Visual saliency for retrieval explanation} highlights image regions and textual cues that most influence similarity scores, revealing the model's decision process.  
(4) \textbf{Prompt enhancement and token attribution} link ranking changes to individual words and offer targeted, interpretable suggestions for query refinement.  
(5) \textbf{Positioning of our work} shows how the system unifies and extends these strands to deliver an explainable and interactive CIR workflow.

\subsection{Visual Analytics for Retrieval Systems}

Recent work in retrieval has increasingly emphasized interactive visual analytics to reveal ranking behavior, query formulation effects, and feedback loops. Early systems such as RankExplorer \cite{shi2012rankexplorer} visualized how rankings evolve over time, while Colorslope \cite{wang2023colorslope} extended this concept with slope graphs and heatmaps to capture dense ranking trajectories. These tools laid the foundation for systems that help users interpret the dynamic behavior of rankers in response to changing queries.

Some systems have explored prompt-based interaction at a higher level. PromptMagician \cite{feng2023promptmagician} supports keyword recommendation and multilevel prompt editing for text-to-image generation, offering a practical interface for experimenting with compositional queries.

Another important direction is counterfactual analysis. CREDENCE \cite{rorseth2023credence} introduces query and document edits to explain ranking shifts, extended by recent frameworks for editable search explanations \cite{xu2023cfe2, chandna2024counterfactual}. In image retrieval, even a single well-chosen keyword has been shown to improve CLIP-based results significantly, without requiring retraining \cite{nara2024cliprf}.

These developments illustrate a shift away from static evaluation toward interactive systems that help users analyze ranking dynamics, explore prompt effects, and understand retrieval behavior-principles that our system adopts and extends.

\subsection{Dimensionality Reduction and Embedding Visualization}
\label{sec:DR_and_emb_vis}

Visualizing high-dimensional data in a meaningful way requires dimensionality reduction techniques that preserve the structure of the original space. Classical methods such as Principal Component Analysis (PCA) \cite{Jolliffe2002Principal} and Isometric Mapping (Isomap) \cite{tenenbaum2000} reduce dimensionality by preserving variance or geodesic distances, respectively. More recent methods such as t-distributed Stochastic Neighbor Embedding (t-SNE) and UMAP emphasize local neighborhood preservation and scalability, making them suitable for large-scale vision-language embeddings \cite{JMLR:v9:vandermaaten08a, mcinnes2020umapuniformmanifoldapproximation}.  
Other techniques including Triplet Map (TriMap) \cite{amid2021}, which uses triplet constraints, and Pairwise Controlled Manifold Approximation and Projection (PaCMAP) \cite{wang2021}, which balances global and local structure through staged optimization, have expanded the design space for projection-based analysis. These approaches have been applied in multimedia tasks, ranging from interactive music exploration \cite{Tovstogan2022Visualization} and fast image annotation \cite{luus19}, to clustering validation in dense text retrieval \cite{liu2022dimensionreductionefficientdense}.  
In our system, we adopt this strategy by using an adjusted UMAP projection to visualize the embedding space in a single, interpretable view that supports retrieval-driven diagnostics.

Building on these dimensionality reduction methods, embedding visualization tools allow users to explore and compare model representations interactively. TensorBoard's Embedding Projector introduced interactive PCA, t-SNE and UMAP plots with brushing and label coloring, providing one of the first practical interfaces for high-dimensional embedding analysis \cite{smilkov2016embeddingprojectorinteractivevisualization}.  
EmbeddingVis expanded this functionality to support side-by-side model comparison \cite{li2018embeddingvisvisualanalyticsapproach}, while WizMap scaled these interactions to millions of points using quadtree summaries and map-style navigation \cite{wang2023wizmapscalableinteractivevisualization}.  
Despite their utility, most of these tools remain model-agnostic and task-agnostic, treating each embedding as an isolated point. They lack retrieval-specific signals such as rank position, prompt evolution, and ideal-image anchors, all of which are integrated into our system \cite{va_embeddings, zahalka2014}.

\subsection{Visual Saliency for Retrieval Explanation}
\label{sec:visual_sal}

Visual saliency helps users interpret why a retrieval model ranks certain images as relevant, by highlighting the most influential regions or features. Similarity-based methods such as \cite{dong2019sbsm} perturb candidate images to identify areas that drive similarity scores, while interactive frameworks like \cite{vasu2021xmir} integrate these maps into feedback loops that enable real-time prompt or crop adjustments. Gradient-based techniques extend saliency to multimodal retrieval: Grad-ECLIP \cite{zhao2025gradeclip} adapts Grad-CAM \cite{Selvaraju_2019} to CLIP by back-propagating through both image and text towers to produce joint heatmaps and token-level gradients, and SANE \cite{plummer2019sane} links patch-level saliency to semantic attributes for improved interpretability. In the context of CIR, where queries combine visual and textual cues, such techniques help users understand how models align the modalities. Our system incorporates and adapts Grad-ECLIP to support post-retrieval explanation: after each search, a heatmap is overlaid on the reference or any retrieved image, connecting salient regions to influential prompt tokens and guiding targeted prompt enhancement.

\subsection{Prompt Enhancement and Token Attribution}

\label{sec:2_4}

Early counterfactual studies showed that small query edits can reveal why certain items appear at incorrect ranks \cite{chandna2024counterfactual}. PromptBench \cite{zhu2024promptbenchunifiedlibraryevaluation} and Counterfact \cite{chen2022greasegeneratefactualcounterfactual} extend this idea to language models by attributing rank shifts to individual tokens instead of entire prompts. 

In the context of multimodal models, gradient-based methods like Grad-ECLIP, SANE, and PromptExp compute signed influence scores for each sub-word across CLIP, image-text, and LLM tasks respectively \cite{zhao2025gradeclip, plummer2019sane, dong2024promptexp}. These studies show that token-level signals more effectively reveal issues such as hallucinated attributes, compared to prompt-level saliency.

While visual summarization tools like RankExplorer \cite{shi2012rankexplorer} and Colorslope \cite{wang2023colorslope} visualize rank shifts under prompt edits, they do not link changes to token-level causes. We adopt their color-encoded Rank-$\Delta$ design, but couple every row with a bar chart of per-token gradients, closing the causality loop between what moved and which word caused the move.

By combining token attribution with rank-change visualization, our system extends explanation techniques originally developed for text-only models to the multimodal setting of CIR. This integration bridges counterfactual query editing with saliency-based interpretation, enabling users to trace ranking changes back to specific words and their corresponding visual evidence.

\subsection{Positioning of our work}

Existing CIR models \cite{agnolucci2024isearleimprovingtextualinversion, efthymiadis2024composedimageretrievaltrainingfree} often focus on isolated components, overlooking the need for an integrated approach. Visual embedding tools typically ignore retrieval context and prompt semantics, while ranking interfaces emphasize accuracy without offering interpretability. Dimensionality reduction methods remain model-agnostic, providing limited insight into retrieval behavior, and prompt enhancement tools usually support only a single modality with minimal explanation. 

Our system addresses these gaps by unifying retrieval ranks, prompt evolution, and token attribution in a single analytics platform, enabling direct comparison across CIR models. To our knowledge, it is the first system to jointly support analysis of embedding space, query dynamics, and model behavior.

\section{Design Rationale \& Goals}
\label{sec:probdef}
The design of InfoCIR is rooted in the need to understand complex model behavior when queries are dynamically changing during the retrieval process. Rather than treating the retrieval as a static operation, our system views it as an iterative dialogue where the user refines multimodal inputs to align the model's internal representation with their specific intent. This section formalizes the retrieval and enhancement tasks and outlines the goals that guide our system's interactive capabilities.

\subsection{Formal Task}

\textbf{Composed Image Retrieval} (CIR) targets the problem of retrieving images that resemble a given reference image but differ according to a textual modification. Unlike traditional image retrieval, CIR requires the system to interpret a natural language transformation and apply it to the visual input, fusing both modalities into a single, semantically meaningful query. This process often involves mapping the reference image into the natural language embedding space, a process known as textual inversion, to allow for a seamless fusion with the text prompt.

Formally, given:
\begin{itemize}[nosep]
    \item a reference image $I_{\text{ref}} \in \mathcal{D}$ (query image),
    \item a natural language modifier $T$ (relative caption),
    \item a database $\mathcal{D}$ of images with precomputed embeddings,
\end{itemize}

the goal is to retrieve a ranked list of the top-$k$ target images that are similar to $I_{\text{ref}}$, except for the modifications described by $T$. This is operationalized as a nearest-neighbor search in a joint multimodal embedding space:

\[
\bigl(I_{(1)},\dots,I_{(k)}\bigr)
\;=\;
\operatorname*{arg\,sort}_{I_j\in\mathcal{D}}
\;\mathrm{sim}\!\Bigl(
            f_{\!c}(I_{\text{ref}},T),
            f_{\!i}(I_j)
          \Bigr),
\]

\noindent
where $f_{\!c}$ is a composition function that fuses the image and text into a query embedding, and $f_{\!i}$ is a fixed encoder for the database images.

\textbf{Prompt enhancement} is the task of rewriting the original text modifier $T$ so that the composed query embeds more accurately the user's intent. After the initial search, the user selects one or more ideal images $I^{*}$ from the top-$k$ results. A language-model-based editor $\mathcal{R}$ then produces a small set of alternative prompts conditioned on the visual context of the reference image $I_{\text{ref}}$,  the original text modifier $T$, and the semantic content of the chosen ideal image(s):

\[
T' = \mathcal{R}\!\bigl(I_{\text{ref}},\,T,\,I^{*}\bigr).
\]

Each candidate prompt $T'$ is re-evaluated by the retrieval engine to determine its impact on the ranking of $I^*$ relative to the original query. This single-pass rewrite aims to boost retrieval quality by better aligning the textual modifier with the visual features of the intended target without requiring model retraining.

\subsection{Interaction Analysis}
\label{sec:interaction_analysis}

To motivate the design of our visual analytics system for CIR, we analyze how users reason while iteratively composing and refining multimodal queries. Rather than issuing a single query and accepting the result, users engage in a continuous sensemaking loop, modeled after Pike et al.’s science of interaction \cite{interaction}, where they must diagnose retrieval failures, compare potential solutions, and enhance their queries to improve outcomes. Within this loop, each interaction functions as a micro-scale hypothesis test that updates the user's mental model of how the system processes multimodal inputs. This structured ``diagnose-compare-enhance'' workflow ensures that interaction leads to both better retrieval results and a deeper understanding of model behavior.

Throughout this process, users pursue a set of recurring high-level interaction intents that structure how CIR results are analyzed. These intents describe users’ macro-level objectives and reflect the cognitive stages of sensemaking described by Pike et al. We characterize these objectives within the three main elements of our proposed workflow:

\begin{itemize}[itemsep=4pt, parsep=0pt, topsep=2pt] 
    \item Diagnose: Users examine why the system behaves as observed to identify retrieval failures. This involves assessing the semantic coherence of the results, identifying structural outliers in the distribution, and detecting over-represented concepts that may be biasing the retrieval.
    
    \item Compare: Users evaluate different states of the retrieval engine to determine which formulation is most effective. This includes switching between prompt variants to observe which produces more coherent semantic neighborhoods and comparing the positional shifts of target images in the embedding space.
    
    \item Enhance: Users iteratively modify the composed prompt to better express the target concept. Each revision represents a new hypothesis that is validated by observing rank changes of target images and leveraging automated suggestions to refine the search focus.
\end{itemize}
These elements are tightly coupled and typically occur within a single sensemaking episode, with users moving fluidly between diagnosis, comparison, and enhancement as they iteratively improve CIR outcomes.

\subsection{Design Goals}
Building on the ``diagnose-compare-enhance'' workflow defined above, we formulate three design goals to operationalize these cognitive stages into specific system capabilities. These goals guide the interface design, ensuring that every visual component directly supports a distinct phase of the user's reasoning process.

\begin{description}[leftmargin=1.55cm, style=sameline, labelsep=0.4em]

\item[DG1 Diagnose] Support rapid assessment and diagnosis of retrieval behavior. The system must provide interpretable visual signals regarding both global semantic structure and local model decisions, enabling users to quickly judge retrieval relevance and pinpoint the causes of reasoning breakdowns.

\item[DG2 Compare] Enable comparison across prompt variants and retrieval outcomes. The system should explicitly visualize shifts in ranking dynamics and semantic proximity between queries, allowing users to contrast outcomes and determine which formulation best captures the intended concept.

\item[DG3 Enhance] Support iterative prompt enhancement. The system must facilitate rapid experimentation by providing intelligent query guidance and immediate feedback on refinements, enabling users to efficiently test new hypotheses without blind trial-and-error.

\end{description}

\section{Methodology}

\begin{figure*}[t!]
    \centering
    \includegraphics[width=1\linewidth]{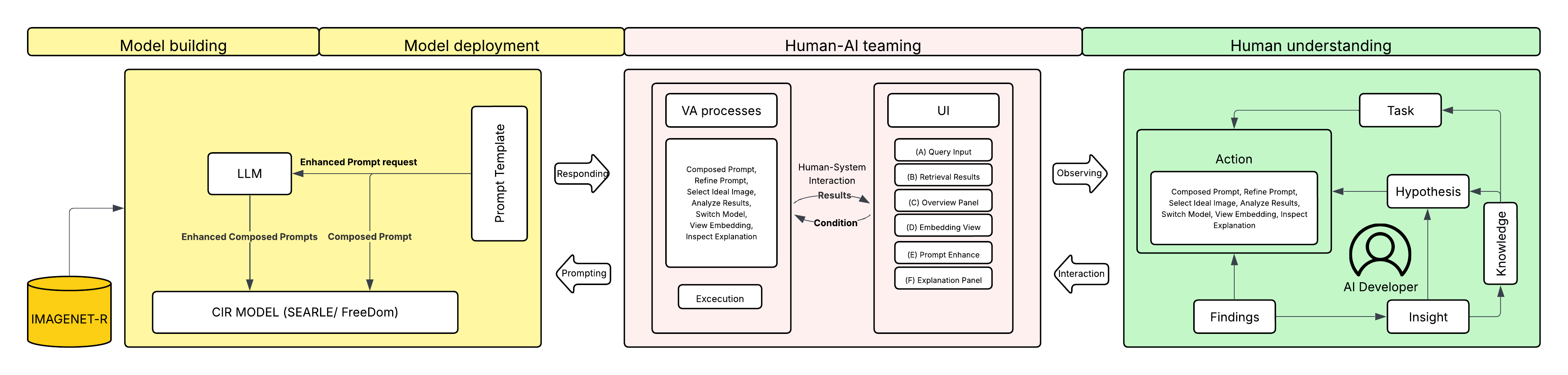}
    \caption{The architecture links prompt composition, embedding-based retrieval, and interactive visualization across UI panels (A-F), enabling users to enhance prompts, interpret outputs, and derive insights. Following Wang et al.~\cite{WANG2025100748}, we integrate semantic fusion (CLIP space), visual integration (saliency and token overlays), and cross-view linking (synchronized views).}

    \label{fig:schema}
\end{figure*}

The proposed interactive multimedia analytics framework for Composed Image Retrieval (CIR) combines multimodal deep learning with coordinated visual representations to support prompt refinement and model introspection. It centers around user feedback, interpretable visualizations and prompt enhancement strategies, leveraging large language models (LLMs). The system follows a modular pipeline architecture, shown in \cref{fig:schema}, inspired by the work of Worring et al. \cite{worring2025multimediaanalyticsmodelfoundation}, with distinct stages for input, model selection, retrieval, projection, visualization and prompt enhancement. The methodological components of the system are described below.

\subsection{System Overview}

The pipeline begins with a user-provided input query consisting of a reference image and an accompanying natural language prompt. It integrates the SEARLE \cite{agnolucci2024isearleimprovingtextualinversion} CIR model, which uses CLIP to compute joint image-text embeddings for comparative retrieval and visualization workflows.

These embeddings are used to rank images in a target database by similarity. The top-$k$ most similar results are visualized using a suite of coordinated views including a 2D embedding view, image gallery, class distribution histogram, and semantic word cloud. An integrated explainable prompt enhancement module uses an LLM to generate alternative text variants, re-ranks them against a user-chosen ideal image, and instantly visualizes their impact through saliency maps, compact Rank-$\Delta$ heatmaps and token-influence summaries, giving users a rapid, transparent loop for refining their queries.

The system is implemented using the Plotly and Dash frameworks, which support highly interactive, web-based visualizations with seamless integration of Python-based machine learning components, as illustrated in the system overview image (\cref{fig:teaser}).

\subsection{Input and Model Selection}

The system accepts two primary inputs: (1) a reference image which is uploaded and (2) a textual prompt describing the desired transformation or characteristics. 
A third, optional (but strongly recommended) input is the ideal image(s): from the initial top-$k$ list the user selects one or more images that align with their intent. These images, become the fixed anchor, against which all prompt variants are evaluated, enabling fine-grained counterfactual diagnostics described in \cref{sec:methodology_prompt_enhancement}.

For the retrieval back-end, InfoCIR implements SEARLE, a zero-shot model that performs textual inversion by transforming the reference image into a pseudo-token within CLIP's text embedding space. This architecture enables the compositional fusion of the image and prompt into a joint query vector. The system then calculates cosine similarities between this vector and pre-computed database embeddings to generate the final image rankings.

\subsection{Retrieval and Similarity Computation}

Given the composed query embedding, the system ranks all database images using cosine similarity. The top-$k$ images are selected for further visualization, forming a ranked result set. Alongside the images, similarity scores and, when available, associated class labels or tags are returned for further analysis.

\subsection{Dimensionality Reduction}

To facilitate qualitative, topological inspection of the embedding structure, the high-dimensional vectors of the top-$k$ results are projected to a two-dimensional space. This projection is intended to support diagnosis through visual neighborhood validation, rather than precise metric reasoning over embedding distances.

In this domain, UMAP and t-SNE stand out as the standard techniques for high-dimensional visualization. We employ UMAP due to three workflow-driven requirements:

\begin{enumerate}
    \item \textbf{Parametric Stability (Out-of-Sample Extension):} 
    Our system embeds new prompts into a fixed reference map in real time. UMAP’s parametric formulation supports stable transform operations on unseen data \cite{Sainburg}, whereas standard t-SNE is non-parametric and typically requires re-running the projection.

    \item \textbf{Global structure preservation:} 
    Like t-SNE, UMAP emphasizes local neighborhood preservation, which is essential for identifying semantically related groups of embeddings \cite{mcinnes2020umapuniformmanifoldapproximation}. However, UMAP typically produces more coherent coarse-grained groups of clusters \cite{chang2025surveypotentialdimensionalityreduction}.

    \item \textbf{Cluster Density for Rapid Semantic Verification:} 
    UMAP tends to produce denser and more compact clusters with clearer inter-class separation \cite{kobak_dense}. This directly supports our DG1 of enabling users to rapidly verify whether a prompt has landed within the intended semantic category.
\end{enumerate}

The resulting two-dimensional coordinates are used to position images in the Embedding View.

\begin{figure}[t!]
    \centering
    \includegraphics[width=1\linewidth]{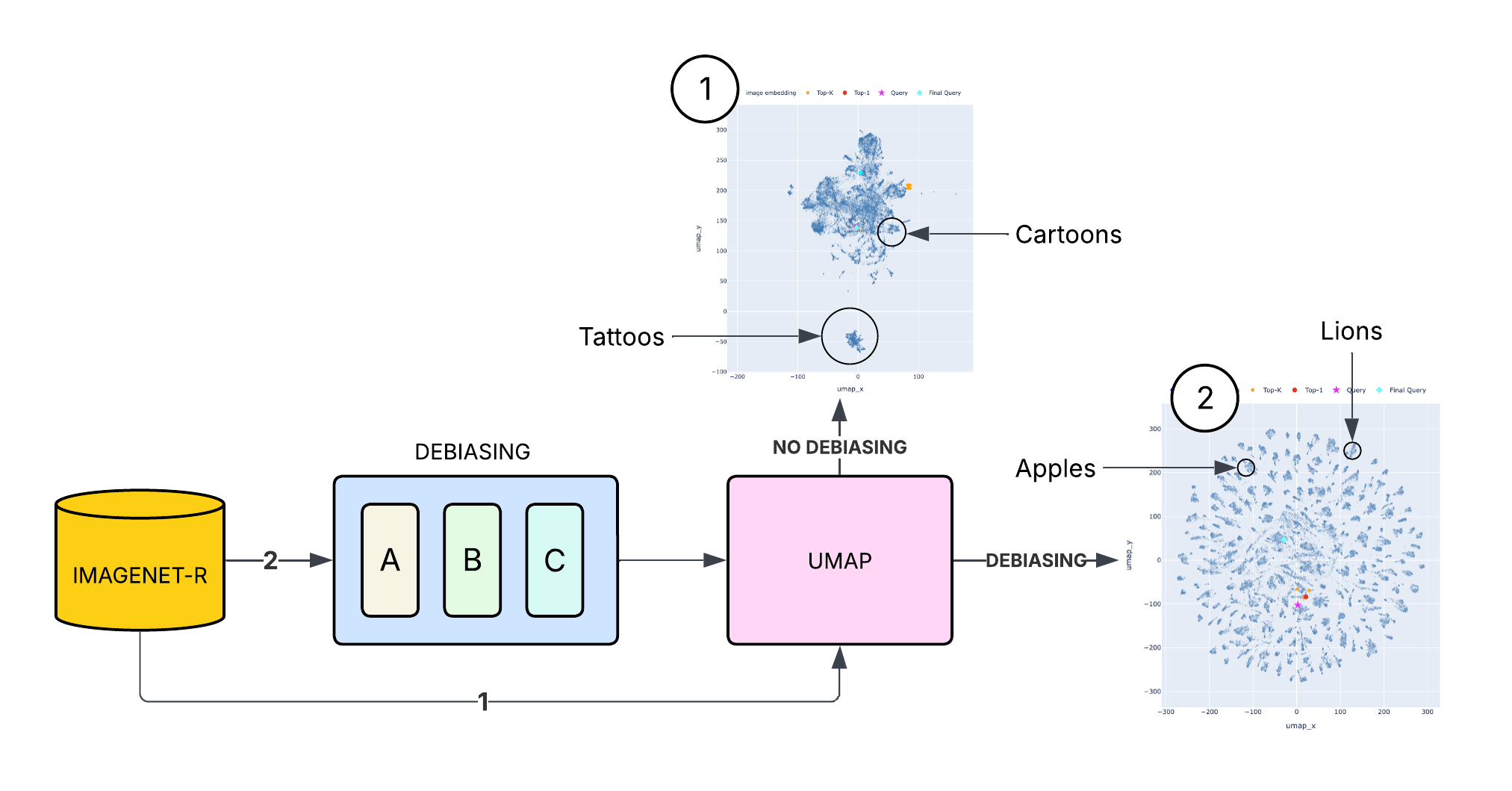}
    \caption{%
        Conceptual overview of the Class Projection Preservation Pipeline. ImageNet-R embeddings are passed through a three-stage filtering process: (A) Style Debiasing, (B) Contrastive Debiasing, and (C) Alternative Projection via ICA, prior to the final UMAP visualization. Compared to the unprocessed projection (1), the processed version (2) exhibits improved class separability and reduced influence of style artifacts, aiding visual verification of class membership.
    }
    \label{fig:Umap_schema}
\end{figure}

To ensure the visualization supports semantic debugging, we apply a Class Projection Preservation Pipeline (Fig.~\ref{fig:Umap_schema}) prior to the final projection. Raw CLIP embeddings often exhibit significant overlap between classes due to stylistic or background features. We explicitly frame this pipeline not as model training or adaptation, but as a heuristic, task-driven semantic pre-alignment strategy applied before visualization.

The pipeline consists of three sequential filtering steps:
(A) \textbf{Style Debiasing} via PCA-based filtering to retain dimensions with higher semantic discriminability;
(B) \textbf{Contrastive Debiasing} \cite{ghosh2020superviseddimensionalityreductionvisualization}, which encourages embeddings to align more closely with their class prototypes;
and (C) \textbf{Independent Component Analysis (ICA)} \cite{poczos2015ica} to separate semantic signals from stylistic noise.

This heuristic pipeline aims to improve class-level separability so that the resulting visualization reflects user-relevant semantic categories rather than irrelevant stylistic variation. Finally, we underline that the Panel D is intended for qualitative, topological inspection rather than precise metric reasoning. Distances and directions in the two-dimensional projection should not be interpreted as faithful representations of semantic magnitude or linear embedding trajectories.

\subsection{Interactive Visualization}

The retrieved and projected images are rendered in six main panels (A–F) as illustrated in \cref{fig:teaser}, providing multiple coordinated views to support the diagnose-compare-enhance workflow and fulfill the design goals:
\begin{itemize}
    \item \textbf{Composed Image Retrieval Panel (A):} Allows users to input a reference image and a text prompt while selecting the number of images $k$ to be retrieved.
    \item \textbf{Query Results Panel (B):} Displays the retrieved images in an Image Gallery sorted by similarity to the query. This view supports rapid inspection of ranked outputs (DG1) and is synchronized with the Embedding View.
    \item \textbf{Histogram / Word Cloud Panel (C):} Includes a class-frequency Histogram ($C_1$) to analyze category dominance and a Word Cloud ($C_2$) that summarizes semantic trends within the current top-$k$. These widgets help users diagnose over-represented concepts or bias in the results, explicitly addressing DG1.
    \item \textbf{Embedding View (D):} The central panel shows a 2D UMAP projection of the dataset. Spatial proximity reflects coarse semantic neighborhood relationships rather than precise similarity magnitudes, allowing users to observe class-level clustering, separation, or outliers (DG1). Interactive selection reveals metadata associated with the respective projection.
    \item \textbf{Prompt Enhancement Panel (E):} Proposes alternative prompt variants $T'$ generated by an LLM, conditioned on user-selected ideal images. Users can compare these variants interactively to observe their impact on retrieval performance, facilitating comparison (DG2) and iterative refinement (DG3).
    \item \textbf{Explanation Panel (F):} Provides multi-level attribution to expose the causes of model behavior, supporting diagnostic reasoning (DG1). This includes:
    \begin{itemize}
        \item \textbf{Saliency Map (F1):} Overlays gradient-based visual saliency to highlight influential image regions, linking visual evidence to textual input and assisting users in understanding model focus.
        \item \textbf{Token Attribution Chart (F2):} Quantifies how strongly each token in the prompt contributes to the retrieval of the ideal image.
        \item \textbf{Rank-$\Delta$ Heatmap (F3):} Visualizes how each variant affects the position of the originally retrieved top-$k$ images. Cell colors show rank changes (green for upward moves, red for drops), allowing users to enhance the search (DG3) by identifying the most effective prompts through direct comparison (DG2).
    \end{itemize}
\end{itemize}
These panels are linked to support multi-faceted exploration. Any interaction in one view, such as clicking an image in the Query Results Panel, automatically updates the associated analytical and explanatory views.

\subsection{Explainable Prompt Enhancement}
\label{sec:methodology_prompt_enhancement}

The prompt enhancement module in InfoCIR is a user-guided implementation of the refinement function $\mathcal{R}$. A specific challenge in this workflow is the ``cold start'' problem, where the initial query fails to retrieve any relevant images in the top-$k$ to serve as feedback anchors. InfoCIR addresses this by utilizing the manual Composed Image Retrieval Panel (Panel A) as a primary pivoting tool: in zero-hit scenarios, users first adjust the initial modifier $T$ to shift the search into a relevant semantic neighborhood. Once a viable target appears in the top-$k$, the user selects one or more ideal target images $I^*$. These chosen images serve as the anchors for enhancement: the system leverages the open-weight mistralai/Mistral-7B-Instruct-v0.2 large language model (LLM) to generate a set of alternative prompts $T'$ grounded on the content of $I_{\text{ref}}$, $T$, and the visual features of $I^*$. For each proposed variant $T'$, the system re-ranks the results and visualizes the rank change of $I^*$ via an Ideal-anchored Rank-$\Delta$ heatmap. This heatmap provides immediate feedback on how each new prompt affects the target's position, helping users compare prompt efficacy at a glance. Because the heatmap fixes the initial top-$k$ as a stable reference set, it visualizes local re-ordering to preserve visual continuity across variants.

\begin{figure}[t!]
    \centering
    \includegraphics[width=1\linewidth]{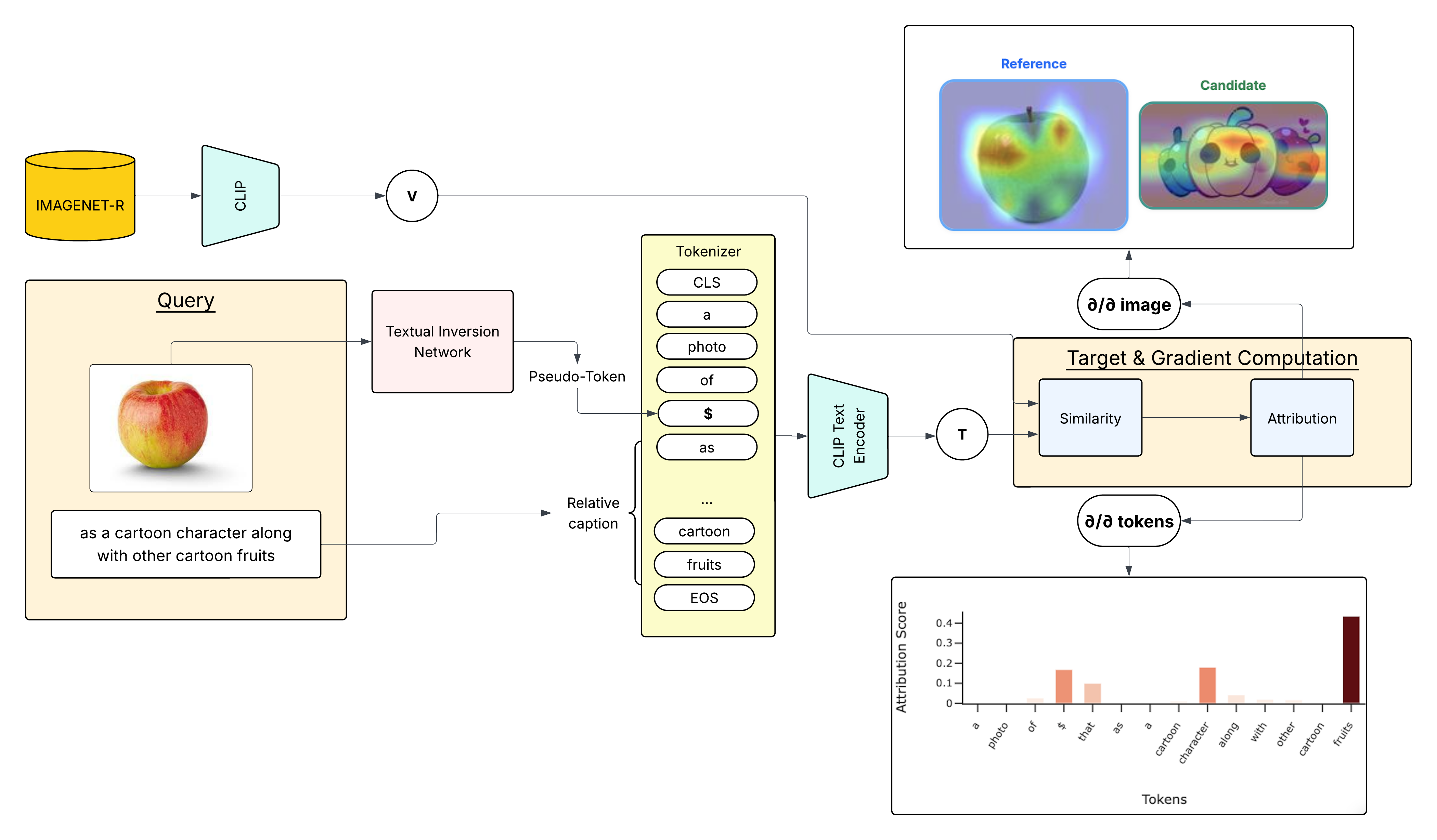}
    \caption{%
    Explanation pipeline for visual and textual attribution in CIR. The system accepts a reference image \( I_r \) and a relative caption \( C \), combining them into the prompt ``a photo of \( \varphi(I_r) \) that \( C \)'', where \( \varphi(I_r) \) is a pseudo-token generated via a textual inversion network. This prompt is tokenized and encoded by CLIP's text encoder to produce a text embedding \( T \). Candidate images from the dataset (e.g., ImageNet-R) are encoded into image embeddings \( V \), and retrieval is based on cosine similarity \( s = \cos(V, T) \). To generate explanations, the similarity score \( s \) is differentiated with respect to either the image or the text tokens. The resulting gradients yield visual saliency maps (top right) and token attribution scores (bottom right), helping users understand which parts of the image or prompt influenced the retrieval result most.%
    }
    \label{fig:saliency_token}
\end{figure}

Along with the prompt suggestions, the module provides tools to diagnose why certain prompts perform better. A token attribution chart (using gradient-based analysis \cite{dong2024promptexp}) highlights which words in $T'$ have the strongest impact on the similarity score $s = \mathrm{sim}(f_c(I_{\text{ref}}, T'), f_i(I^*))$. Technically, we compute the gradient of $s$ with respect to each input token. The magnitude of these gradients indicates each word's contribution to the match, which is visualized as a bar chart to guide users toward high-influence terms. This offers insight into the model's reasoning about the text query and suggests which keywords might be emphasized or adjusted in further refinements (see \cref{fig:saliency_token}).

Complementing the token-level explanation, a saliency map is generated to show which regions of $I_{\text{ref}}$ or $I^*$ are most influential. We employ gradient-based visual explanation techniques, and more specifically implement from scratch an adaptation of Grad-ECLIP \cite{zhao2025gradeclip} for our CLIP-based multimodal encoder (SEARLE), to attribute the output of $f_c$ back to image pixels. By back-propagating the gradient of $s$ through the model's image tower, we produce a heatmap overlay denoting areas crucial to the query-result similarity. The saliency highlights enable users to see which parts of the image the model ``looked at'' when matching the query, linking concrete visual evidence to the semantic query feedback (see \cref{fig:saliency_token}).

Together, these explainable prompt enhancement features put the user in control of an interactive feedback loop. The LLM-generated variants $T'$ (with their rank-change indicators) and the attribution visualizations (token bars and saliency maps) collectively, allow users to steer the search with greater transparency and confidence. This approach aligns with emerging interactive explanation paradigms that emphasize human-in-the-loop refinement and deeper model transparency \cite{rorseth2023credence}, ensuring that prompt adjustments are both informed by the model's inner workings and grounded on the user's intent, ultimately fulfilling DG3 by lowering the barrier to effective experimentation.

\subsection{System Modularity and Extensibility}

The system is designed with modularity in mind. Retrieval models can be easily swapped or added as long as they provide image and text embeddings in the CLIP embedding space. Visualization modules are decoupled and configurable, supporting the addition of new views or encoding strategies. This flexibility allows for experimentation with retrieval algorithms, datasets and user interaction paradigms.

\subsection{Usage Scenario}
\label{sec:usage_scenario}

To demonstrate how InfoCIR supports interactive refinement in CIR, consider a user uploading an image of a green apple and entering the prompt ``a red apple''. The goal is to retrieve semantically relevant images that preserve the object identity (apple) while modifying its attribute (color).

Upon querying, the SEARLE retriever returns a result set that includes both green and red apples, along with unrelated fruits. In the UMAP projection, results are scattered, with no clear cluster for red apples. The user opens the attribution panel, which shows that the term ``red'' contributes weakly to the ranking, suggesting a style bias in the CLIP representation.

To correct this, the user selects one or more retrieved images that match their intent (e.g., any clearly deep-red apple) and opens the Prompt Enhancement module. This component suggests semantically guided prompt variants, such as ``ripe apple'', ``crimson apple'', and ``dark red fruit'', each grounded on corpus embeddings and visual similarity.

After choosing ``crimson apple'', the updated retrieval results appear in a compact cluster of red apples in the embedding space. The selected image jumps from rank 12 to rank 2, as shown in the Rank-$\Delta$ heatmap, and the attribution weights now reflect stronger activation on ``red'' and relevant image regions.

This scenario illustrates how InfoCIR enables users to debug and steer the retrieval process through prompt adaptation and embedding inspection, leading to semantically faithful results that align with the composed intent.

\section{Evaluation}

Our evaluation strategy focuses on the Prompt Enhancement workflow as the primary unit of analysis. While InfoCIR comprises six coordinated views (A-F), prompt enhancement acts as the ``keystone'' task that necessitates the synthesis of information from the entire dashboard. To successfully refine a query, a user must first diagnose the initial failure (using the Embedding View and Saliency Maps), formulate a hypothesis (using the LLM suggestions), and verify the outcome (using the Rank-$\Delta$ heatmap). Consequently, we argue that measuring performance on prompt enhancement serves as a holistic proxy for the utility of the supporting diagnostic components.

We conducted a two-part evaluation to assess how InfoCIR supports users in efficiently retrieving target images and understanding model behavior through its interactive analytical widgets. The study combined quantitative performance tasks with qualitative feedback analysis to examine both objective efficiency and subjective user experience. Our main hypothesis was that InfoCIR's interactive explanatory features, particularly the Prompt Enhancement Panel, would improve retrieval performance, user confidence, and understanding of image-text relationships compared to a baseline system without such support.

This study adhered to established ethical research standards. All participants were informed about the study and provided written consent. Participation was voluntary, no identifying data were collected, and all responses were anonymized and stored securely. In line with GDPR regulations, the data will be permanently deleted after the project is completed.

\subsection{Quantitative Study: Performance and Retrieval Efficiency}

\subsubsection{Procedure}

We recruited eight Master's students in Artificial Intelligence from the University of Amsterdam, all with a background in computer science. Participants had previously worked with web development, design, or information retrieval tools, and had experience searching for and retrieving images as part of past professional tasks. Henceforth, we will refer to the participants as P1 through P8.

\begin{figure}[ht!]
    \centering
    \setlength{\tabcolsep}{6pt}
    \renewcommand{\arraystretch}{1.1}
    \begin{tabular}{cc}
        \includegraphics[width=0.24\columnwidth]{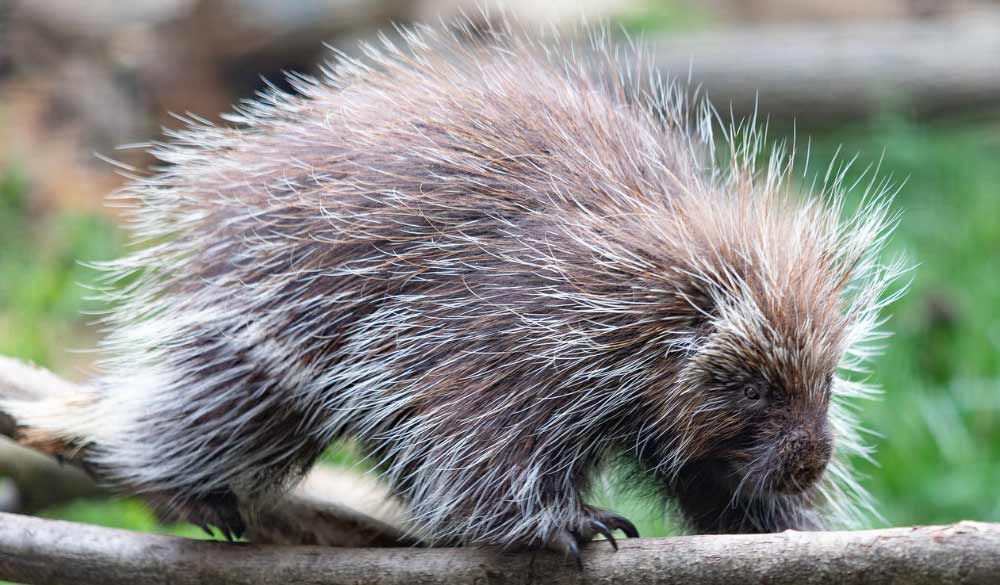} &
        \includegraphics[width=0.20\columnwidth]{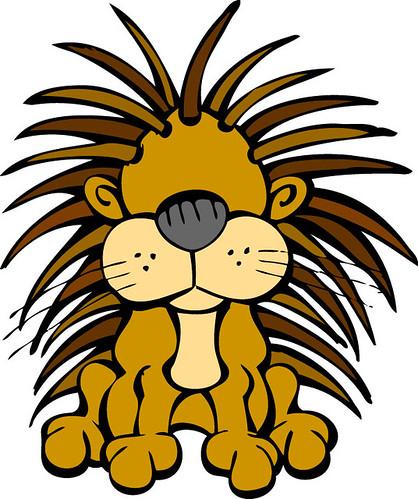} \\
        \textbf{(a) Reference Task A} & 
        \textbf{(b) Target Task A} \\
        \includegraphics[width=0.24\columnwidth]{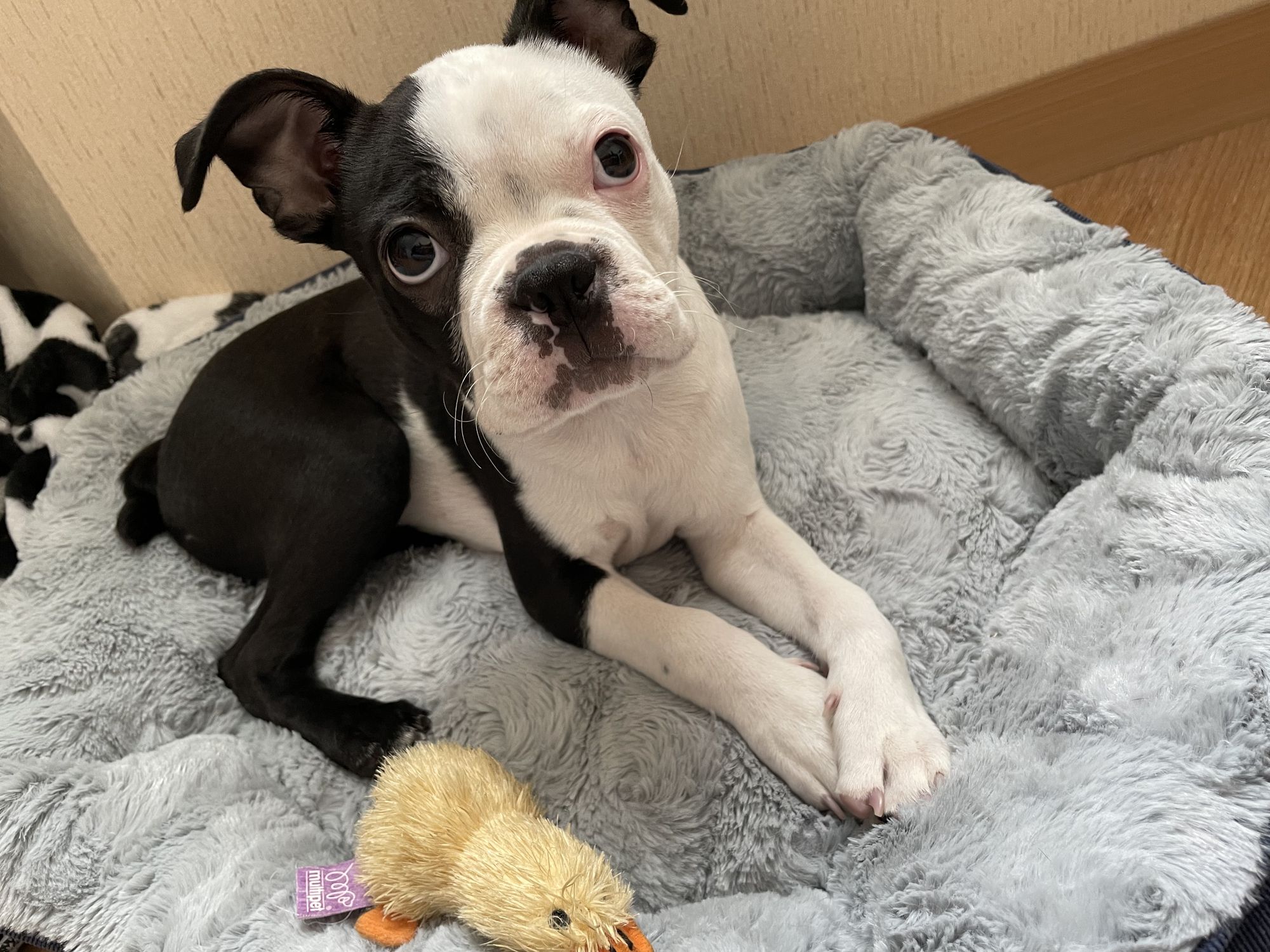} &
        \includegraphics[width=0.24\columnwidth]{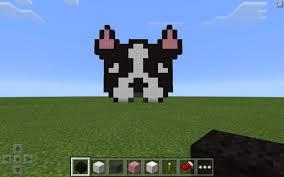} \\
        \textbf{(c) Reference Task B} & 
        \textbf{(d) Target Task B} \\
    \end{tabular}
    \caption{Reference and target image pairs used in the quantitative evaluation. Task A (porcupine) and Task B (Boston Terrier) each required retrieving a stylized version of a realistic reference image. Participants iteratively refined a text prompt until the target appeared within the top-3 retrieved results.}
    \label{fig:tasks}
\end{figure}

For the dataset, we used ImageNet-R to simulate realistic retrieval scenarios with similar difficulty and visual abstraction. The reference images for Task A and Task B were obtained from publicly available online sources\footnote{Reference image for Task A, retrieved from: https://brandywinezoo.org/wp-content/uploads/2024/10/north-american-porcupine-brandywine-zoo-2023.jpg} \footnote{Reference image for Task B, retrieved from: https://dogtime.com/wp-content/uploads/sites/12/2023/09/GettyImages-1500664833.jpg}. Two image retrieval tasks (Task A and Task B) were designed to have comparable semantic and perceptual difficulty. In both tasks, participants viewed a realistic reference image and attempted to retrieve a stylized (cartoon-like) version of the same object category. Task A involved a \textbf{porcupine} and Task B a \textbf{Boston Terrier} (see \cref{fig:tasks}). Both categories were selected for their similar visual complexity and distinct, recognizable features.

\textbf{Design and Procedure} 
We compared participants' performance while using two different versions of the tool, one basic and one enhanced, to see how the widgets affect the image retrieval task.
\begin{enumerate}[leftmargin=2em, itemsep=1pt, parsep=0pt, topsep=2pt]
    \item \textbf{Scenario 1 - Baseline:} Participants used only the Composed Image Retrieval and Query Results panels, with all InfoCIR's widgets hidden.
    \item \textbf{Scenario 2 - InfoCIR:} Participants used the full InfoCIR interface, including the Prompt Enhancement Panel, Saliency Map, Histogram/Word Cloud, Embedding View, Token Attribution, and Rank-$\Delta$ heatmap.
\end{enumerate}

To avoid order bias, participants were split into two groups that completed the same tasks in reversed order, ensuring task sequence did not influence performance.

\begin{itemize}[leftmargin=2em, itemsep=1pt, parsep=0pt, topsep=2pt]
    \item \textbf{Group A:} Task A (Baseline) $\rightarrow$ Task B (InfoCIR)
    \item \textbf{Group B:} Task B (Baseline) $\rightarrow$ Task A (InfoCIR)
\end{itemize}

Each task required participants to refine a text prompt until the target appeared in the top-3 retrieved images, with a 5-minute limit. Each query returned ten images. Sessions were timed, and two experimenters recorded results. Performance was evaluated using:

\begin{enumerate}[leftmargin=2em, itemsep=1pt, parsep=0pt, topsep=2pt]
    \item Success Rate: proportion of participants who placed the target in the top-3.
    \item Average Time on Task: mean time to successful retrieval.
    \item Number of Unique Queries: average number of prompt submissions.
\end{enumerate}

\subsubsection{Quantitative Findings}

As summarized in \cref{tab:participantresults}, participants using the full InfoCIR interface showed markedly improved retrieval performance compared to the baseline. The success rate increased from 37.50\% to 87.50\%, average completion time decreased from 277 seconds to 133 seconds, and the average number of unique queries dropped from 7.4 to 3.3. One participant (P6) did not succeed in either condition, as the initial provided prompts were overly generic and the target did not appear within the top ten retrieved images, leaving no meaningful feedback to guide refinement and highlighting that effective use of InfoCIR still requires purposeful prompt adjustment.

\begin{table*}[tb]
  \caption{Participant performance across the two retrieval scenarios. Scenario~A is the baseline (manual prompt refinement). Scenario~B is the InfoCIR-assisted condition (all analytical widgets and the Prompt Enhancement Panel enabled). Success indicates whether the target image appeared in the top-3 retrieval results. For Scenario~B, each prompt refinement action was also counted as a query. Averages are across all 8 participants.}
  \label{tab:participantresults}
  \scriptsize
  \centering
  \begin{tabu}{l c *{6}{r}}
    \toprule
    \multicolumn{2}{l}{} & \multicolumn{3}{c}{\textbf{Scenario A (Baseline)}} & \multicolumn{3}{c}{\textbf{Scenario B (InfoCIR-Assisted)}} \\
    \cmidrule(lr){3-5}\cmidrule(lr){6-8}
    \textbf{ID} & \textbf{Group} & \textbf{Time (s)} & \textbf{Queries} & \textbf{Success Rate (\%)} & \textbf{Time (s)} & \textbf{Queries} & \textbf{Success Rate (\%)} \\
    \midrule
    P1 & A & 243 & 6  & 100.00 & 103 & 3 & 100.00 \\
    P2 & A & 300 & 9  &   0.00 &  87 & 2 & 100.00 \\
    P3 & A & 300 & 7  &   0.00 & 134 & 3 & 100.00 \\
    P4 & A & 281 & 5  & 100.00 &  91 & 3 & 100.00 \\
    P5 & B & 300 & 11 &   0.00 & 149 & 4 & 100.00 \\
    P6 & B & 300 & 8  &   0.00 & 300 & 6 &   0.00 \\
    P7 & B & 194 & 4  & 100.00 &  55 & 2 & 100.00 \\
    P8 & B & 300 & 9  &   0.00 & 142 & 3 & 100.00 \\
    \midrule
    \textbf{Average} & -- & \textbf{277} & \textbf{7.4} & \textbf{37.50} & \textbf{133} & \textbf{3.3} & \textbf{87.50} \\
    \bottomrule
  \end{tabu}
\end{table*}

These results indicate that InfoCIR's widgets accelerated convergence by guiding users toward more effective, category-based language. The metrics confirm the fulfillment of the design goals: the reduced time and query counts demonstrate support for rapid experimentation (DG3), while the high success rate shows that users successfully diagnosed and corrected retrieval failures (DG1).

\subsection{Qualitative Study: User Experience and Interpretability}

\subsubsection{Procedure}

After completing both retrieval tasks, participants filled out a post-study questionnaire containing five-point Likert scale evaluations of each InfoCIR component. A short discussion followed in which participants described their workflow, what they found intuitive, and where they encountered difficulties. The goal of this qualitative phase was to understand how users experienced the interface while performing retrieval, which components they relied on most, and how the system supported their reasoning.

\subsubsection{Qualitative Findings}

Participants generally agreed that the enhanced interface improved both retrieval efficiency and their understanding of how the system processes prompts. However, several participants (P1, P3, P5, P8) noted an initial period of orientation, during which they explored the various components before converging on a smaller set of features that they relied on consistently. This indicates a learning curve, especially for components intended to aid interpretability rather than immediate task execution.

The \textbf{Prompt Enhancement Panel} was consistently identified as the most useful feature for improving retrieval. The majority of participants rated its suggestions as clear and relevant, and 62.50\% of participants selected it as the most helpful widget overall. Participants claimed that ``It made the task easier'' (P4) and ``I do not need to manually craft a strong initial prompt in order to retrieve the target image within the top-3 results'' (P7). Moreover, participants indicated that using the panel taught them how to phrase category-based prompts more effectively rather than relying on descriptive attributes of the target image (figure available in supplementary material), validating its role in supporting iterative prompt enhancement (DG3).

The \textbf{Histogram and Word Cloud} helped participants identify more effective vocabulary by showing which concepts the model associated with their current prompt. Participants reported that it supported the shift from descriptive phrasing toward category based terms. For example, P2 explained that it helped change ``small black and white dog'' to ``Boston Terrier'', which directly improved retrieval, and P5 described it as a useful guide for discovering the model's preferred terminology (figure available in supplementary material). By exposing these semantic mismatches, these widgets successfully fulfilled the goal of supporting rapid diagnosis (DG1). However, participants also noted that while the widget aided understanding, it contributed more to overall orientation than to fast prompt refinement during the task.

Both the \textbf{Embedding View} and the \textbf{Saliency Map} contributed meaningfully to model interpretability and post hoc reasoning rather than immediate retrieval decisions. Participants generally understood the Embedding View and used it primarily to observe how the system groups images and organizes similarity relationships. Several participants reported that it helped them grasp the model's internal structure but not in a way that directly supported fast prompt refinement. P5 noted that it ``explains the structure, but I cannot use it quickly''. Concurrently, the Saliency Map assisted users by verifying which image regions influenced retrieval and was often employed as a correctness check (figure available in supplementary material). P5 described it as useful for confirming whether the system was focusing on the main object, aligning with the diagnostic requirements of DG1. However, participants commonly noted a temporal cost, commenting that it required more time and attention than the task allowed. P3 stated that it ``takes too much time to deal with'', and P4 mentioned that although it clarified model attention, it did not directly help with prompt adjustment.

The \textbf{Token Attribution} and \textbf{Rank-$\Delta$ Heatmap} were used less frequently during the time-constrained retrieval tasks, as many participants found them slower to interpret and therefore difficult to apply to fast prompt adjustments. However, these widgets supported deeper analysis for participants who adopted a more exploratory approach. P3 and P8, in particular, spent additional time examining these visualizations during Scenario B and reported that they helped them understand how subtle phrasing changes influenced ranking outcomes (figure available in supplementary material). This suggests that these widgets are valuable for detailed interpretability and comparison (DG2) when users have sufficient time to explore them.

\section{Discussion and Future Work}

The development of InfoCIR highlights the growing need for explainable, interactive systems in the domain of Composed Image Retrieval (CIR). While recent CIR models show promising zero-shot and training-free retrieval capabilities \cite{agnolucci2024isearleimprovingtextualinversion, efthymiadis2024composedimageretrievaltrainingfree}, their behavior often remains opaque to users, particularly when multimodal queries (image + text) yield unexpected results. Our system addresses this challenge by providing a tightly integrated workflow, through which users can trace retrieval dynamics through interpretable and visual means.

A key design insight was the benefit of combining retrieval results with both the overall structure of the embedding space and detailed feedback on individual prompt tokens. Embedding projections via UMAP support exploration of neighborhood coherence and cluster structure, while token attribution and saliency maps provide insights on the internal reasoning behind model decisions. The connected and interactive visual views allow users to detect and act on misalignments, such as when a prompt term appears dominant despite being semantically not relevant, making it easier to quickly form and test new ideas.

Another important contribution is the idea of ideal-guided prompt enhancement. By allowing users to fix one or more target images as benchmarks, InfoCIR turns prompt enhancement into a constrained optimization task visible through the Rank-$\Delta$ heatmap. This approach offers immediate, localized feedback on prompt effectiveness, addressing a longstanding gap in model debugging workflows, where prompt quality is typically judged only by final retrieval rank without insight into causal dynamics \cite{chandna2025counterfactualexplanationframeworkretrieval, lauro2025raglaginteractivedebugging}.

However, limitations remain:

\begin{itemize}[itemsep=1pt, parsep=0pt, topsep=0pt]
    \item \textbf{Limitations of Embedding Visualization}
    Projecting high-dimensional CLIP embeddings into 2D inherently introduces metric distortion, meaning that linear vector arithmetic valid in the original space does not strictly hold in the projection. Consequently, the Embedding View is intended for visual cluster verification, confirming topological containment within a semantic cluster, rather than as a precise metric map for trajectory analysis. While our pipeline facilitates this task by creating distinct clusters for rapid visual inspection, we acknowledge the inherent trade-offs in manifold approximation. Future work will explore alternative techniques such as PaCMAP and TriMap \cite{wang2021,amid2021,fu2025mpadnewdimensionreductionmethod} to potentially achieve a better balance between local fidelity and global structural constraints.

    \item \textbf{Cold Start Vulnerability:} The automated prompt enhancement relies on the user selecting at least one relevant ``ideal'' image from the initial results. This creates a dependency on the initial query performance. If the zero-shot retrieval fails to return any relevant targets (a ``zero-hit'' scenario), the user must rely on manual iteration to pivot the search into a relevant semantic neighborhood before the automated enhancement tools can be engaged.

    \item \textbf{Generalization vs. Overfitting:} By anchoring prompt refinement to specific user-selected images, there is a risk that the generated prompts may overfit to the visual features of the chosen anchors rather than generalizing to the broader semantic category. A prompt that successfully retrieves the specific anchor might inadvertently exclude other relevant variations of the target class.

    \item \textbf{Model Actionability:} While InfoCIR effectively diagnoses retrieval failures, the insights are primarily actionable for prompt engineering and query weighting rather than model correction. In the era of large foundation models, developers rarely have the capacity to retrain the heavy backbone (e.g., SEARLE or CLIP) based on granular, instance-level observations exposed by the tool.

    \item \textbf{LLM Heuristics and Verification:} Prompt suggestions are generated by an open-weight LLM and remain heuristic; they do not guarantee retrieval improvement and may occasionally hallucinate attributes. We mitigate this via the Rank-$\Delta$ heatmap, which acts as a verification loop, allowing users to empirically test and discard ineffective variants. Future work could incorporate structured prompts (e.g., scene graphs \cite{yin2024sgnavonline3dscene}) or multi-turn refinement \cite{ye2025promptalchemyautomaticprompt} to improve generation stability \cite{app15095198}.

    \item \textbf{Attribution Fidelity:} Token attribution charts depend on gradient approximations that may not fully capture the complex attention mechanisms in transformer-based architectures \cite{hatefi2024pruningexplainingrevisitedoptimizing}.

    \item \textbf{Local vs. Global Explainability:} Attribution-based explanations offer only local insight into specific query-image pairs. Extending the system to support counterfactual visual generation (e.g., what-if edits to the query image) or concept-level abstractions could provide global interpretability and support model debugging at a higher level \cite{ALFEO2023107550}.

    \item \textbf{Feedback Loop Integration:} Current refinement consists of a single LLM-based rewrite guided only by the user-selected ideal images. Saliency maps and token-attribution bars serve purely as explanatory overlays and are not fed back into the rewrite function. Future iterations could let explicit user annotations, such as region sketches, custom attention masks, or reinforcement-guided scoring, directly influence the prompt generator, providing finer control and richer engagement.
\end{itemize}

Despite these constraints, InfoCIR demonstrates that aligning retrieval outcomes with human-centered reasoning is both feasible and impactful. The system advances CIR tooling beyond black-box metrics toward interactive, explanation-rich workflows that better support model development, evaluation and refinement.

\section{Conclusion}

We presented InfoCIR, a visual analytics system designed to enhance transparency, interpretability and refinement in Composed Image Retrieval (CIR). By integrating retrieval ranking, embedding projection and visualizations, prompt enhancement and multi-level explanation within a coordinated and interactive interface, our system addresses key gaps in CIR interpretability and supports developers in exploring, understanding, improving and validating retrieval model behavior with greater confidence.

Our contributions include an explainable prompt-enhancement pipeline that combines LLM-generated prompt variants with gradient-based attribution visualizations, a coordinated environment for inspecting embedding dynamics, and a modular architecture that can accommodate new CIR models with minimal effort. Through guided interface interactions, users can rapidly assess prompt impact, explore the embedding landscape, and pinpoint failure cases or performance bottlenecks.

In contrast to existing tools that isolate embedding inspection or prompt engineering, InfoCIR combines both, supporting a new paradigm of retrieval-centered multimodal analysis.

Future work can extend model support, scalability and personalization, further bridging the gap between opaque model internals and user-centered retrieval workflows. As the field of multimodal AI continues to evolve, we see InfoCIR as both a diagnostic toolkit and a framework for future systems that prioritize human-guided refinement, transparency and trust in retrieval pipelines.

\section*{Supplemental Materials}
\label{sec:supplemental_materials}

Supplemental materials have been submitted and include (1) An appendix with the UMAP-adjusted Results and the Widgets during the Evaluation Process, (2) a demo of the InfoCIR system, and (3) a hyperlink to the GitHub repository containing the source code of the InfoCIR system.

\newpage
\bibliographystyle{abbrv-doi}

\bibliography{bibliography}

@misc{baldrati2023zeroshotcomposedimageretrieval,
      title={Zero-Shot Composed Image Retrieval with Textual Inversion}, 
      author={Alberto Baldrati and Lorenzo Agnolucci and Marco Bertini and Alberto Del Bimbo},
      year={2023},
      eprint={2303.15247},
      archivePrefix={arXiv},
      primaryClass={cs.CV},
      url={https://arxiv.org/abs/2303.15247}, 
}

@article{interaction,
author = {Pike, William A. and Stasko, John and Chang, Remco and O'Connell, Theresa A.},
title = {The science of interaction},
year = {2009},
issue_date = {December 2009},
publisher = {Palgrave Macmillan},
volume = {8},
number = {4},
issn = {1473-8716},
url = {https://doi.org/10.1057/ivs.2009.22},
doi = {10.1057/ivs.2009.22},
journal = {Information Visualization},
month = dec,
pages = {263–274},
numpages = {12}
}

@misc{agnolucci2024isearleimprovingtextualinversion,
      title={iSEARLE: Improving Textual Inversion for Zero-Shot Composed Image Retrieval}, 
      author={Lorenzo Agnolucci and Alberto Baldrati and Marco Bertini and Alberto Del Bimbo},
      year={2024},
      eprint={2405.02951},
      archivePrefix={arXiv},
      primaryClass={cs.CV},
      url={https://arxiv.org/abs/2405.02951}, 
}

@misc{efthymiadis2024composedimageretrievaltrainingfree,
      title={Composed Image Retrieval for Training-Free Domain Conversion}, 
      author={Nikos Efthymiadis and Bill Psomas and Zakaria Laskar and Konstantinos Karantzalos and Yannis Avrithis and Ondřej Chum and Giorgos Tolias},
      year={2024},
      eprint={2412.03297},
      archivePrefix={arXiv},
      primaryClass={cs.CV},
      url={https://arxiv.org/abs/2412.03297}, 
}

@misc{smilkov2016embeddingprojectorinteractivevisualization,
      title={Embedding Projector: Interactive Visualization and Interpretation of Embeddings}, 
      author={Daniel Smilkov and Nikhil Thorat and Charles Nicholson and Emily Reif and Fernanda B. Viégas and Martin Wattenberg},
      year={2016},
      eprint={1611.05469},
      archivePrefix={arXiv},
      primaryClass={stat.ML},
      url={https://arxiv.org/abs/1611.05469}, 
}

@misc{li2018embeddingvisvisualanalyticsapproach,
      title={EmbeddingVis: A Visual Analytics Approach to Comparative Network Embedding Inspection}, 
      author={Quan Li and Kristanto Sean Njotoprawiro and Hammad Haleem and Qiaoan Chen and Chris Yi and Xiaojuan Ma},
      year={2018},
      eprint={1808.09074},
      archivePrefix={arXiv},
      primaryClass={cs.HC},
      url={https://arxiv.org/abs/1808.09074}, 
}

@misc{wang2023wizmapscalableinteractivevisualization,
      title={WizMap: Scalable Interactive Visualization for Exploring Large Machine Learning Embeddings}, 
      author={Zijie J. Wang and Fred Hohman and Duen Horng Chau},
      year={2023},
      eprint={2306.09328},
      archivePrefix={arXiv},
      primaryClass={cs.LG},
      url={https://arxiv.org/abs/2306.09328}, 
}

@article{va_embeddings,
author = {Huang, Zeyang and Witschard, Daniel and Kucher, Kostiantyn and Kerren, Andreas},
year = {2023},
month = {06},
pages = {539-571},
title = {VA + Embeddings STAR: A State‐of‐the‐Art Report on the Use of Embeddings in Visual Analytics},
volume = {42},
journal = {Computer Graphics Forum},
doi = {10.1111/cgf.14859}
}

@article{shi2012rankexplorer,
  title     = {RankExplorer: Visualization of Ranking Changes in Large Time Series Data},
  author    = {Conglei Shi and Weiwei Cui and Shixia Liu and Panpan Xu and Wei Chen and Huamin Qu},
  journal   = {IEEE Transactions on Visualization and Computer Graphics},
  volume    = {18},
  number    = {12},
  pages     = {2669--2678},
  year      = {2012},
  doi       = {10.1109/TVCG.2012.253}
}

@article{wang2023colorslope,
  title     = {Colorslope: A Balanced Visualization of Overview and Details on Ranks over Time},
  author    = {Hao Wang and Xingyu Jiang and Apurva Nagarajan and Xiaolei Guo and Lu Ding and Dayu Wan and Junhan Zhao and Yingjie Chen},
  journal   = {Visual Intelligence},
  volume    = {1},
  number    = {7},
  pages     = {1--18},
  year      = {2023},
  doi       = {10.1007/s44267-023-00008-9}
}

@article{feng2023promptmagician,
  title     = {PromptMagician: Interactive Prompt Engineering for Text to Image Creation},
  author    = {Yingchaojie Feng and Xingbo Wang and Kam Kwai Wong and Sijia Wang and Yuhong Lu and Minfeng Zhu and Baicheng Wang and Wei Chen},
  journal   = {IEEE Transactions on Visualization and Computer Graphics},
  volume    = {30},
  number    = {1},
  pages     = {295--305},
  year      = {2024},
  doi       = {10.1109/TVCG.2023.3327168}
}

@article{dong2024promptexp,
  title     = {PromptExp: Multi Granularity Prompt Explanation of Large Language Models},
  author    = {Ximing Dong and Shaowei Wang and Dayi Lin and Gopi Krishnan Rajbahadur and Boquan Zhou and Shichao Liu and Ahmed E. Hassan},
  journal   = {CoRR},
  volume    = {abs/2410.13073},
  year      = {2024},
  url       = {https://arxiv.org/abs/2410.13073}
}

@article{zhao2025gradeclip,
  title     = {Grad ECLIP: Gradient Based Visual and Textual Explanations for CLIP},
  author    = {Chenyang Zhao and Kun Wang and Janet H. Hsiao and Antoni B. Chan},
  journal   = {CoRR},
  volume    = {abs/2502.18816},
  year      = {2025},
  url       = {https://arxiv.org/abs/2502.18816}
}

@inproceedings{rorseth2023credence,
  title     = {CREDENCE: Counterfactual Explanations for Document Ranking},
  author    = {Joel Rorseth and Parke Godfrey and Lukasz Golab and Mehdi Kargar and Divesh Srivastava and Jaroslaw Szlichta},
  booktitle = {Proceedings of the 39th International Conference on Data Engineering},
  year      = {2023},
  url       = {https://arxiv.org/abs/2302.04983}
}

@article{xu2023cfe2,
  title     = {Counterfactual Editing for Search Result Explanation},
  author    = {Zhichao Xu and Hemank Lamba and Qingyao Ai and Joel Tetreault and Alex Jaimes},
  journal   = {CoRR},
  volume    = {abs/2301.10389},
  year      = {2023},
  url       = {https://arxiv.org/abs/2301.10389}
}

@article{chandna2024counterfactual,
  title     = {A Counterfactual Explanation Framework for Retrieval Models},
  author    = {Bhavik Chandna and Procheta Sen},
  journal   = {CoRR},
  volume    = {abs/2409.00860},
  year      = {2024},
  url       = {https://arxiv.org/abs/2409.00860}
}

@article{nara2024cliprf,
  title     = {Revisiting Relevance Feedback for CLIP Based Interactive Image Retrieval},
  author    = {Ryoya Nara and Yu Chieh Lin and Yuji Nozawa and Youyang Ng and Goh Itoh and Osamu Torii and Yusuke Matsui},
  journal   = {CoRR},
  volume    = {abs/2404.16398},
  year      = {2024},
  url       = {https://arxiv.org/abs/2404.16398}
}

@article{tenenbaum2000,
  title     = {A Global Geometric Framework for Nonlinear Dimensionality Reduction},
  author    = {Tenenbaum, Joshua B. and De Silva, Vin and Langford, John C.},
  journal   = {Science},
  volume    = {290},
  number    = {5500},
  pages     = {2319--2323},
  year      = {2000},
  doi       = {10.1126/science.290.5500.2319}
}

@article{JMLR:v9:vandermaaten08a,
  author  = {Laurens van der Maaten and Geoffrey Hinton},
  title   = {Visualizing Data using t-SNE},
  journal = {Journal of Machine Learning Research},
  year    = {2008},
  volume  = {9},
  number  = {86},
  pages   = {2579--2605},
  url     = {http://jmlr.org/papers/v9/vandermaaten08a.html}
}

@misc{mcinnes2020umapuniformmanifoldapproximation,
      title={UMAP: Uniform Manifold Approximation and Projection for Dimension Reduction}, 
      author={Leland McInnes and John Healy and James Melville},
      year={2020},
      eprint={1802.03426},
      archivePrefix={arXiv},
      primaryClass={stat.ML},
      url={https://arxiv.org/abs/1802.03426}, 
}

@misc{amid2021,
      title={TriMap: Large-scale Dimensionality Reduction Using Triplets}, 
      author={Ehsan Amid and Manfred K. Warmuth},
      year={2022},
      eprint={1910.00204},
      archivePrefix={arXiv},
      primaryClass={cs.LG},
      url={https://arxiv.org/abs/1910.00204}, 
}

@misc{wang2021,
      title={Understanding How Dimension Reduction Tools Work: An Empirical Approach to Deciphering t-SNE, UMAP, TriMAP, and PaCMAP for Data Visualization}, 
      author={Yingfan Wang and Haiyang Huang and Cynthia Rudin and Yaron Shaposhnik},
      year={2021},
      eprint={2012.04456},
      archivePrefix={arXiv},
      primaryClass={cs.LG},
      url={https://arxiv.org/abs/2012.04456}, 
}

@inproceedings{dong2019sbsm,
  author    = {Bo Dong and Roddy Collins and Anthony Hoogs},
  title     = {Explainability for Content-Based Image Retrieval},
  booktitle = {Proceedings of the IEEE/CVF Conference on Computer Vision and Pattern Recognition Workshops},
  pages     = {95--98},
  year      = {2019},
  url       = {https://openaccess.thecvf.com/content_CVPRW_2019/papers/ExplainableAI_CUP/Dong_Explainability_for_Content-Based_Image_Retrieval_CVPRW_2019_paper.pdf}
}

@article{vasu2021xmir,
  author  = {Bharath Vasu and Boxiang Hu and Bo Dong and Roddy Collins and Anthony Hoogs},
  title   = {Explainable Interactive Content-Based Image Retrieval},
  journal = {Applied AI Letters},
  volume  = {2},
  number  = {4},
  pages   = {e41},
  year    = {2021},
  doi     = {10.1002/ail2.41},
  url     = {https://onlinelibrary.wiley.com/doi/10.1002/ail2.41}
}

@article{plummer2019sane,
  author  = {Bryan A.\ Plummer and Mariya I.\ Vasileva and Vitali Petsiuk and Kate Saenko and David Forsyth},
  title   = {Why Do These Match? Explaining the Behavior of Image Similarity Models},
  journal = {arXiv},
  year    = {2019},
  archivePrefix = {arXiv},
  eprint  = {1905.10797},
  url     = {https://arxiv.org/abs/1905.10797}
}

@article{Selvaraju_2019,
   title={Grad-CAM: Visual Explanations from Deep Networks via Gradient-Based Localization},
   volume={128},
   ISSN={1573-1405},
   url={http://dx.doi.org/10.1007/s11263-019-01228-7},
   DOI={10.1007/s11263-019-01228-7},
   number={2},
   journal={International Journal of Computer Vision},
   publisher={Springer Science and Business Media LLC},
   author={Selvaraju, Ramprasaath R. and Cogswell, Michael and Das, Abhishek and Vedantam, Ramakrishna and Parikh, Devi and Batra, Dhruv},
   year={2019},
   month=oct, pages={336–359} }

@misc{zhu2024promptbenchunifiedlibraryevaluation,
      title={PromptBench: A Unified Library for Evaluation of Large Language Models}, 
      author={Kaijie Zhu and Qinlin Zhao and Hao Chen and Jindong Wang and Xing Xie},
      year={2024},
      eprint={2312.07910},
      archivePrefix={arXiv},
      primaryClass={cs.AI},
      url={https://arxiv.org/abs/2312.07910}, 
}

@misc{chen2022greasegeneratefactualcounterfactual,
      title={GREASE: Generate Factual and Counterfactual Explanations for GNN-based Recommendations}, 
      author={Ziheng Chen and Fabrizio Silvestri and Jia Wang and Yongfeng Zhang and Zhenhua Huang and Hongshik Ahn and Gabriele Tolomei},
      year={2022},
      eprint={2208.04222},
      archivePrefix={arXiv},
      primaryClass={cs.IR},
      url={https://arxiv.org/abs/2208.04222}, 
}

@inproceedings{Tovstogan2022Visualization,
  author    = {Philip Tovstogan and Xavier Serra and Dmitry Bogdanov},
  title     = {Visualization of Deep Audio Embeddings for Music Exploration and Rediscovery},
  booktitle = {Proceedings of the 19th Sound and Music Computing Conference (SMC)},
  year      = {2022},
  pages     = {493--500},
  address   = {Saint-Étienne, France},
  month     = jun,
  url       = {http://hdl.handle.net/10230/53710},
  note      = {Available from UPF Digital Repository},
}

@inproceedings{luus19,
author = {Luus, Francois and Khan, Naweed and Akhalwaya, Ismail},
title = {Interactive Supervision with t-SNE},
year = {2019},
isbn = {9781450370080},
publisher = {Association for Computing Machinery},
address = {New York, NY, USA},
url = {https://doi.org/10.1145/3360901.3364414},
doi = {10.1145/3360901.3364414},
booktitle = {Proceedings of the 10th International Conference on Knowledge Capture},
pages = {85–92},
numpages = {8},
location = {Marina Del Rey, CA, USA},
series = {K-CAP '19}
}

@misc{liu2022dimensionreductionefficientdense,
      title={Dimension Reduction for Efficient Dense Retrieval via Conditional Autoencoder}, 
      author={Zhenghao Liu and Han Zhang and Chenyan Xiong and Zhiyuan Liu and Yu Gu and Xiaohua Li},
      year={2022},
      eprint={2205.03284},
      archivePrefix={arXiv},
      primaryClass={cs.IR},
      url={https://arxiv.org/abs/2205.03284}, 
}

@INPROCEEDINGS{zahalka2014,
  author={Zahálka, Jan and Worring, Marcel},
  booktitle={2014 IEEE Conference on Visual Analytics Science and Technology (VAST)}, 
  title={Towards interactive, intelligent, and integrated multimedia analytics}, 
  year={2014},
  volume={},
  number={},
  pages={3-12},
  doi={10.1109/VAST.2014.7042476}}

@article{WANG2025100748,
title = {Empowering multimodal analysis with visualization: A survey},
journal = {Computer Science Review},
volume = {57},
pages = {100748},
year = {2025},
issn = {1574-0137},
doi = {https://doi.org/10.1016/j.cosrev.2025.100748},
url = {https://www.sciencedirect.com/science/article/pii/S1574013725000243},
author = {Jiachen Wang and Zikun Deng and Dazhen Deng and Xingbo Wang and Rui Sheng and Yi Cai and Huamin Qu}
}

@misc{chandna2025counterfactualexplanationframeworkretrieval,
      title={A Counterfactual Explanation Framework for Retrieval Models}, 
      author={Bhavik Chandna and Procheta Sen},
      year={2025},
      eprint={2409.00860},
      archivePrefix={arXiv},
      primaryClass={cs.IR},
      url={https://arxiv.org/abs/2409.00860}, 
}

@misc{lauro2025raglaginteractivedebugging,
      title={RAG Without the Lag: Interactive Debugging for Retrieval-Augmented Generation Pipelines}, 
      author={Quentin Romero Lauro and Shreya Shankar and Sepanta Zeighami and Aditya Parameswaran},
      year={2025},
      eprint={2504.13587},
      archivePrefix={arXiv},
      primaryClass={cs.HC},
      url={https://arxiv.org/abs/2504.13587}, 
}

@misc{fu2025mpadnewdimensionreductionmethod,
      title={MPAD: A New Dimension-Reduction Method for Preserving Nearest Neighbors in High-Dimensional Vector Search}, 
      author={Jiuzhou Fu and Dongfang Zhao},
      year={2025},
      eprint={2504.16335},
      archivePrefix={arXiv},
      primaryClass={cs.IR},
      url={https://arxiv.org/abs/2504.16335}, 
}

@misc{hatefi2024pruningexplainingrevisitedoptimizing,
      title={Pruning By Explaining Revisited: Optimizing Attribution Methods to Prune CNNs and Transformers}, 
      author={Sayed Mohammad Vakilzadeh Hatefi and Maximilian Dreyer and Reduan Achtibat and Thomas Wiegand and Wojciech Samek and Sebastian Lapuschkin},
      year={2024},
      eprint={2408.12568},
      archivePrefix={arXiv},
      primaryClass={cs.AI},
      url={https://arxiv.org/abs/2408.12568}, 
}

@misc{yin2024sgnavonline3dscene,
      title={SG-Nav: Online 3D Scene Graph Prompting for LLM-based Zero-shot Object Navigation}, 
      author={Hang Yin and Xiuwei Xu and Zhenyu Wu and Jie Zhou and Jiwen Lu},
      year={2024},
      eprint={2410.08189},
      archivePrefix={arXiv},
      primaryClass={cs.CV},
      url={https://arxiv.org/abs/2410.08189}, 
}

@misc{ye2025promptalchemyautomaticprompt,
      title={Prompt Alchemy: Automatic Prompt Refinement for Enhancing Code Generation}, 
      author={Sixiang Ye and Zeyu Sun and Guoqing Wang and Liwei Guo and Qingyuan Liang and Zheng Li and Yong Liu},
      year={2025},
      eprint={2503.11085},
      archivePrefix={arXiv},
      primaryClass={cs.SE},
      url={https://arxiv.org/abs/2503.11085}, 
}

@Article{app15095198,
AUTHOR = {Choi, Jaekeol},
TITLE = {Efficient Prompt Optimization for Relevance Evaluation via LLM-Based Confusion Matrix Feedback},
JOURNAL = {Applied Sciences},
VOLUME = {15},
YEAR = {2025},
NUMBER = {9},
ARTICLE-NUMBER = {5198},
URL = {https://www.mdpi.com/2076-3417/15/9/5198},
ISSN = {2076-3417},
DOI = {10.3390/app15095198}
}

@article{ALFEO2023107550,
title = {From local counterfactuals to global feature importance: efficient, robust, and model-agnostic explanations for brain connectivity networks},
journal = {Computer Methods and Programs in Biomedicine},
volume = {236},
pages = {107550},
year = {2023},
issn = {0169-2607},
doi = {https://doi.org/10.1016/j.cmpb.2023.107550},
url = {https://www.sciencedirect.com/science/article/pii/S0169260723002158},
author = {Antonio Luca Alfeo and Antonio G. Zippo and Vincenzo Catrambone and Mario G.C.A. Cimino and Nicola Toschi and Gaetano Valenza}
}

@misc{worring2025multimediaanalyticsmodelfoundation,
      title={A Multimedia Analytics Model for the Foundation Model Era}, 
      author={Marcel Worring and Jan Zahálka and Stef van den Elzen and Maximilian T. Fischer and Daniel A. Keim},
      year={2025},
      eprint={2504.06138},
      archivePrefix={arXiv},
      primaryClass={cs.MM},
      url={https://arxiv.org/abs/2504.06138}, 
}

@misc{vo2018composingtextimageimage,
      title={Composing Text and Image for Image Retrieval - An Empirical Odyssey}, 
      author={Nam Vo and Lu Jiang and Chen Sun and Kevin Murphy and Li-Jia Li and Li Fei-Fei and James Hays},
      year={2018},
      eprint={1812.07119},
      archivePrefix={arXiv},
      primaryClass={cs.CV},
      url={https://arxiv.org/abs/1812.07119}, 
}

@misc{song2025comprehensivesurveycomposedimage,
      title={A Comprehensive Survey on Composed Image Retrieval}, 
      author={Xuemeng Song and Haoqiang Lin and Haokun Wen and Bohan Hou and Mingzhu Xu and Liqiang Nie},
      year={2025},
      eprint={2502.18495},
      archivePrefix={arXiv},
      primaryClass={cs.MM},
      url={https://arxiv.org/abs/2502.18495}, 
}

@misc{psomas2024composedimageretrievalremote,
      title={Composed Image Retrieval for Remote Sensing}, 
      author={Bill Psomas and Ioannis Kakogeorgiou and Nikos Efthymiadis and Giorgos Tolias and Ondrej Chum and Yannis Avrithis and Konstantinos Karantzalos},
      year={2024},
      eprint={2405.15587},
      archivePrefix={arXiv},
      primaryClass={cs.CV},
      url={https://arxiv.org/abs/2405.15587}, 
}

@inproceedings{LDRE,
author = {Yang, Zhenyu and Xue, Dizhan and Qian, Shengsheng and Dong, Weiming and Xu, Changsheng},
title = {LDRE: LLM-based Divergent Reasoning and Ensemble for Zero-Shot Composed Image Retrieval},
year = {2024},
isbn = {9798400704314},
publisher = {Association for Computing Machinery},
address = {New York, NY, USA},
url = {https://doi.org/10.1145/3626772.3657740},
doi = {10.1145/3626772.3657740},
booktitle = {Proceedings of the 47th International ACM SIGIR Conference on Research and Development in Information Retrieval},
pages = {80–90},
numpages = {11},
location = {Washington DC, USA},
series = {SIGIR '24}
}

@misc{tang2025missingtargetrelevantinformationprediction,
      title={Missing Target-Relevant Information Prediction with World Model for Accurate Zero-Shot Composed Image Retrieval}, 
      author={Yuanmin Tang and Jing Yu and Keke Gai and Jiamin Zhuang and Gang Xiong and Gaopeng Gou and Qi Wu},
      year={2025},
      eprint={2503.17109},
      archivePrefix={arXiv},
      primaryClass={cs.CV},
      url={https://arxiv.org/abs/2503.17109}, 
}

@misc{radford2021learningtransferablevisualmodels,
      title={Learning Transferable Visual Models From Natural Language Supervision}, 
      author={Alec Radford and Jong Wook Kim and Chris Hallacy and Aditya Ramesh and Gabriel Goh and Sandhini Agarwal and Girish Sastry and Amanda Askell and Pamela Mishkin and Jack Clark and Gretchen Krueger and Ilya Sutskever},
      year={2021},
      eprint={2103.00020},
      archivePrefix={arXiv},
      primaryClass={cs.CV},
      url={https://arxiv.org/abs/2103.00020}, 
}

@book{Jolliffe2002Principal,
  address = {New York},
  author = {Jolliffe, I. T.},
  booktitle = {Principal Component Analysis},
  comment = {(private-note)Full textbook. I have printed some bits; relevant bit for atm sci is {\S}4.3.},
  doi = {10.1007/b98835},
  isbn = {0-387-95442-2},
  publisher = {Springer-Verlag},
  series = {Springer Series in Statistics},
  title = {Principal Component Analysis},
  url = {http://www.springer.com/statistics/statistical+theory+and+methods/book/978-0-387-95442-4},
  year = 2002
}

@misc{ghosh2020superviseddimensionalityreductionvisualization,
      title={Supervised Dimensionality Reduction and Visualization using Centroid-encoder}, 
      author={Tomojit Ghosh and Michael Kirby},
      year={2020},
      eprint={2002.11934},
      archivePrefix={arXiv},
      primaryClass={cs.LG},
      url={https://arxiv.org/abs/2002.11934}, 
}

@misc{poczos2015ica,
  author       = {Barnabás Póczos},
  title        = {Independent Component Analysis},
  howpublished = {\url{https://www.cs.cmu.edu/~bapoczos/Classes/ML10715_2015Fall/slides/ICA.pdf}},
  note         = {Lecture slides, CMU 10-715: Advanced Introduction to Machine Learning},
  year         = {2015},
  institution  = {Carnegie Mellon University},
}

@ARTICLE{sedlmair2012design,
  author={Sedlmair, Michael and Meyer, Miriah and Munzner, Tamara},
  journal={IEEE Transactions on Visualization and Computer Graphics}, 
  title={Design Study Methodology: Reflections from the Trenches and the Stacks}, 
  year={2012},
  volume={18},
  number={12},
  pages={2431-2440},
  doi={10.1109/TVCG.2012.213}}

@article{Sainburg,
author = {Sainburg, Tim and McInnes, Leland and Gentner, Timothy},
year = {2021},
month = {10},
pages = {2881-2907},
title = {Parametric UMAP Embeddings for Representation and Semisupervised Learning},
volume = {33},
journal = {Neural Computation},
doi = {10.1162/neco_a_01434}
}

@article{kobak_dense,
author = {Kobak, Dmitry and Linderman, George},
year = {2021},
month = {02},
pages = {1-2},
title = {Initialization is critical for preserving global data structure in both t-SNE and UMAP},
volume = {39},
journal = {Nature Biotechnology},
doi = {10.1038/s41587-020-00809-z}
}

@misc{chang2025surveypotentialdimensionalityreduction,
      title={A Survey: Potential Dimensionality Reduction Methods}, 
      author={Yuan-chin Ivan Chang},
      year={2025},
      eprint={2502.11036},
      archivePrefix={arXiv},
      primaryClass={stat.OT},
      url={https://arxiv.org/abs/2502.11036}, 
}
\end{document}